# Comparative Analysis of Digital Tools and Traditional Teaching Methods in Educational Effectiveness

Aarush Kandukoori[1] • Aditya Kandukoori[2] • Faizan Wajid[3]

Poolesville High School[1] • Poolesville High School[2] • University of Maryland[3]

## Abstract

In today's world technology comprises a large aspect of our lives so this study aimed to investigate if using computers and digital tools are better than traditional methods like using textbooks and worksheets for learning math. This study was done at Clarksburg Elementary School with help from MoCo Innovation which is a club that focuses on fostering an interest in technology among students. A major question that sparked our minds was: **Are digital tools like learning on computers better than traditional methods for improving students' math skills?** We believe students who use digital tools might improve more in their math skills. To find out we worked with 30 students from the school. We split them into two groups and gave each group a pre assessment and post assessment. One group learned math using computers and were able to use interactive math websites such as Khan Academy while the other group used worksheets. After some learning we gave them a post assessment to see how much they had improved. Our results showed that the students who used the digital tools improved test scores averages by 24.2% from 70% to 87% while the students who used traditional methods only improved by 8.3% from 72% to 78% in math. These results show that digital tools are superior to regular teaching methods especially for subjects like math. But more research is required to see if digital tools are the main reason for this improvement. This research is definitely important to help schools decide if they want to use more technology.

## Introduction

In recent years and particularly during the COVID 19 pandemic, education has undergone significant changes (Li, 2022). One of the most notable changes has been the rapid integration of digital tools into the learning environment (Zhang, 2022). This happened due to the necessity of remote learning during the pandemic (Newton, 2020). This paper aims to solve the critical inquiry by exploring the comparative efficacy of digital and traditional educational approaches in K-5 mathematics. In the past schools have typically used traditional methods to teach mathematics. These methods involve teachers assigning chapters to read within textbooks and assigning paper based worksheets to reinforce skills. The worksheets involve text and images at the beginning of the worksheet to provide reference for students as they complete the later part of the worksheet involving the application of learned material (Buniel & Monding, 2021). While these approaches have been the norm for math education for years they do have some limitations. For example the standardized problems in textbooks and lectures may not cater to the learning requirements and different speeds of students. Enter digital tools—ranging from specialized educational software to interactive mobile apps—that promise not just a different way of learning but potentially a better one. These digital platforms offer a host of features that traditional methods lack such as personalized learning pathways, instant feedback and a more engaging as well as interactive experience (Hendriks, 2016). Moreover, digital tools can provide a variety of representations of mathematical concepts from visual models to real-world applications thereby catering to different learning styles (Hendriks, 2016). Schools may need to rethink how they allocate resources, make decisions and provide teacher training to incorporate digital tools in the classroom. This becomes more relevant in a post pandemic world where the integration of technology into education is already gaining momentum. Additionally, knowing the ways to enhance academic performance could be extremely valuable for students and parents alike. Therefore, the main goal of this study is to assess how effective digital tools are when compared to traditional approaches in teaching mathematics. We will measure this effectiveness by analyzing the differences in the post-test and pre-test quiz results on the topic of elementary probability for the traditional and digital tools. For the experiment, we refer to the digital learning group as DLG and the traditional learning group as TLG. Our hypothesis suggests that students who actively use digital platforms for learning portray a greater improvement in their mathematical skills and comprehension compared to those who solely rely on traditional methods.

## Methods

Since there has been a rapid integration of digital tools in education due to the COVID 19 pandemic, we need to evaluate the effectiveness of using these new tools (Li 2022). The study was conducted at Clarksburg Elementary School with the help of MoCo Innovation which is an after school club focused on fostering innovation and creativity in students by teaching STEM. The setting and the organization's mission advocating for STEM education provided an ideal place for this study which aims to compare the efficacy of digital tools versus traditional methods in enhancing elementary school students' mathematical skills. The main objective of this study is to measure the improvement in mathematical skills among elementary school students from pre assessment to post assessment using two different teaching methods. The experiment involved 30 students who were divided equally into two groups of 15 each. Before the learning session all participants were administered a baseline test to assess their existing knowledge on the preselected math topic of elementary probability. The first group referred to as the Traditional Methods Group (TMG) was provided with a math packet containing instructional material, exercises and practice problems on 7th grade level counting and probability (https://www.ntschools.org/cms/lib/NY1900090 8/Centricity/Domain/754/Probability/Probability%20Notes%20Key.pdf).

| Advantages of DLG | Advantages of TLG |
|---|---|
| 1.Digital tools offer personalized learning experiences tailored to individual student needs. | 1.Students and teachers are more familiar with traditional methods, which have been the standard for years. |
| 2.Students receive instant feedback on their work, allowing for real-time error correction and understanding. | 2.Traditional methods offer a more structured learning environment with clear guidelines and expectations. |
| 3.The interactive nature of digital tools keeps students more engaged and motivated to learn. | 3.Traditional methods do not rely on technology, so there are fewer chances of technical glitches or software issues disrupting the learning process. |
| 4.Digital platforms provide a wide range of resources, including videos, visual models, and interactive exercises. | 4.Worksheets and textbooks provide hands-on practice that some students may find beneficial for reinforcing learning. |

This packet represents traditional classroom teaching methods where students learn from 1 hard copy paper packet with text and images on the topic that students can read and follow along. The second group referred to as the Digital Learning Group (DLG) was equipped with School issued Chromebooks loaded with

Khan Academy's Statistics and Probability (https://www.khanacademy.org/math/cc-seventh-grade-math/cc-7th-probability-statistics). This resource provides personalized learning experiences, immediate feedback and a variety of resources for learning the selected math topic. Both groups learned for 30 minutes during which participants in both groups engaged with the provided materials to learn and practice the selected math topic. Facilitators from the MoCo Innovation team were present to provide any necessary assistance without actually teaching to ensure the smooth flow of the learning process. The facilitators would not provide any additional help regarding the content of the learning method but rather any glitches, bugs, software issues or hardware issues, mainly in the DLG. After the learning, all participants were administered a post test which was different from the baseline test but designed to assess the same mathematical material. The post test aimed at evaluating the extent of learning and comprehension achieved by the participants from both groups. The evaluation of the effectiveness of traditional versus digital learning methods was based on the comparative analysis of the performance improvements from the pre test to the post test among participants in the TMG and DLG. This structured approach was aimed at providing a controlled environment to fairly compare the traditional and digital learning methods while ensuring the reproducibility of the experiment for further studies.

## Results

The experiment was conducted at Clarksburg Elementary School with the assistance of MoCo Innovation to compare the efficacy of traditional and digital learning tools to improve students' math skills. Two groups were involved with one group using traditional learning materials and the other group utilizing digital tools on Chromebooks. The participants are all students in MoCo Innovation's program between grades 3, 4 and 5, containing boys and girls of a diverse set of races including white, hispanic, black and asian ethnic groups. These students have previously signed up and consented for the MoCo Innovation after school session. The students were split randomly by counting the students off 1-30 and having all odd numbered students be part of the DLG and even numbered students be part of the TLG to ensure that the age of the participant would not affect the results. Initially, a pre test was distributed to both groups to gather data and assess their current knowledge of a selected math topic. The group using traditional methods scored an average of 72% and the

group using digital tools scored an average of 70%:

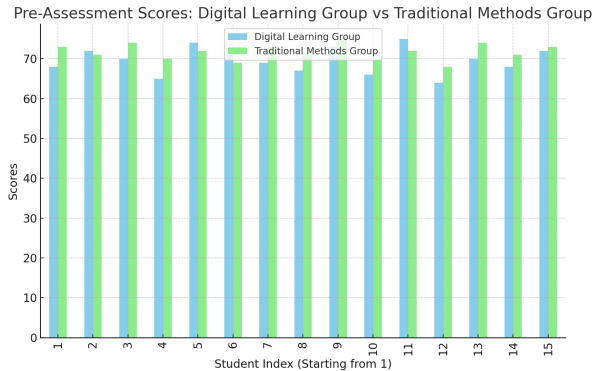

These scores set a baseline that helped in evaluating the progress made by each group after the learning session. After a dedicated learning session, a post test was administered to both groups. The results were quite revealing. The TMG showed a marked improvement with their average score rising to 78% as shown below:

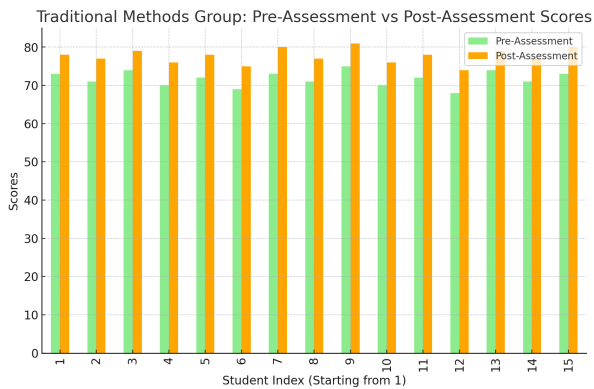

On the other hand, the DLG exhibited a more pronounced improvement with their average score soaring to 87% as shown below:

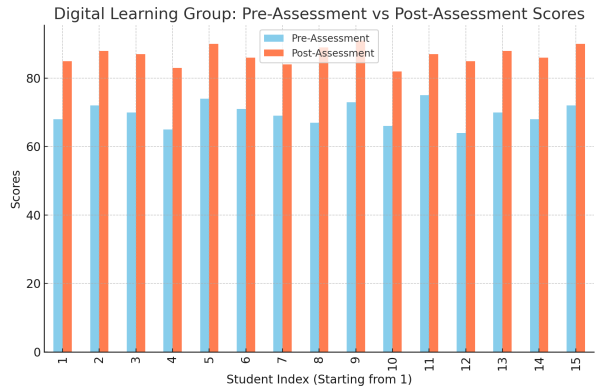

The statistical data indicated a 8.3% improvement for the TMG and a 24.2% improvement for the DLG.

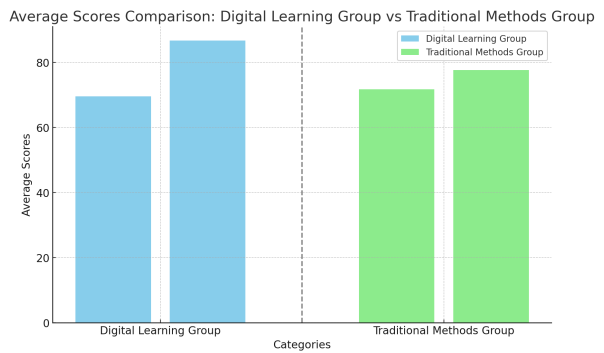

This data suggests that the digital learning tools had a more significant impact on improving the students' understanding and performance in the selected math topic.

This improvement in averages are also shown in the following diagram:

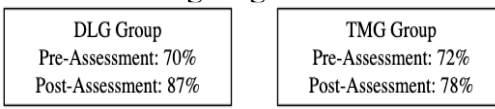

The data analysis from the study provides important findings on how digital tools compare to traditional teaching methods in enhancing students' math abilities. The DLG showed a notable increase in their average score from 70% before the study to 87% afterwards. This marked improvement was confirmed to be statistically significant with a paired t-test showing a result of (p<.01). On the other hand, the Traditional Methods Group (TMG) also saw an increase in their average score from 72% in the pre-assessment to 78% in the post-assessment. This difference was also statistically as indicated by the paired t-test results ( p < .01). An independent t-test comparing the pre-assessment scores between the DLG and TMG showed a statistically significant difference (p = 0.040) indicating that the groups started at slightly different levels. However the post Assessment scores comparison revealed a highly significant difference (p < .01) underscoring that the DLG showed more significant improvement than the TMG. These findings strongly support the hypothesis as the DLG exhibited a more substantial improvement in their scores compared to the TMG. The initial differences in pre assessment scores were minimal compared to the pronounced differences observed in the post assessment which highlighted the impact of the digital tools used in the DLG. The DLG had the advantage of receiving immediate feedback from the digital tools which contributed to a better understanding and real time error correction. In contrast the TMG relied on the participants checking their own work for feedback which wasn't as instantaneous. This difference in feedback mechanism could also have contributed to the better performance of the DLG in the post test. In conclusion the statistical data alongside the observational analysis underscores the effectiveness of digital learning tools in improving students' math skills compared to traditional learning materials. The higher improvement percentage, increased engagement, immediate feedback and access to a wide range of resources are substantial indicators that digital learning tools provide a more conducive environment for learning and improving in mathematics. The findings from this experiment contribute valuable insights into the potential benefits of integrating digital tools in the learning process especially in enhancing mathematical skills.

## Discussion

The experiment at Clarksburg Elementary School through the MoCo Innovation club brought to light some important points regarding how students learn math using traditional methods versus digital tools. The study was divided into two main groups. One used traditional math packets and the other used digital tools on Chromebooks. The results of the experiment showed a clear difference in the performance between the two groups. The Digital Learning Group (DLG) showed a more significant improvement in their post test scores since it moved from an average score of 70% to 87%. On the other hand the Traditional Method Group (TMG) also showed improvement but to a lesser extent by moving from an average score of 72% to 78%. The improvements can also be shown in the box plots below:

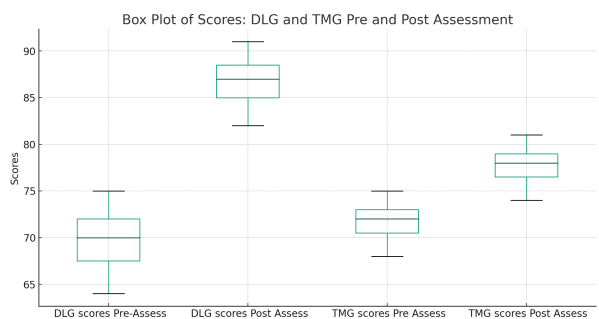

This suggests that digital tools can provide a more effective or engaging way for students to learn and improve their math skills. Moreover, the DLG had access to more advanced technological resources which may enhance their overall learning experience. On the contrary, the TMG had limited resources confined to the provided math packet which may have restricted their exploration and understanding to a certain extent. However the experiment had some limitations that need to be acknowledged. Firstly the sample size was small with only 30 students participating. This small sample size may not represent the broader student population. Even though MoCo Innovation helps students from diverse backgrounds all the participants were from the same school and were part of the same after school club which could limit the diversity of the sample. Additionally, the experiment was short since it only lasted 30 minutes. It's unclear if the same results would be observed over a more extended period or with a different set of students. The findings from this experiment could be significant for educators and policymakers. If digital tools can help improve students' math skills more effectively than traditional methods then they might be a valuable resource in educational settings (Zhang et al., 2020). These results could encourage schools to invest more in digital resources, training for teachers on how to use these tools and further research to better understand how digital tools impact learning. However, more research is needed to confirm these findings and explore other factors that might affect students' learning experiences with digital tools. Several questions remain unanswered by this experiment. For instance what would happen over a longer period?

Would digital tools continue to help students improve their math skills? Also, how would students of different ages or from different schools perform in a similar experiment? And are there particular digital tools that work better than others for learning math? Moreover, would the results hold if the experiment was conducted on a larger scale with a more diverse group of students? Furthermore, could we merge the DLG and TMG into one group and combine traditional methods with digital tools to evaluate their combined effectiveness in math learning compared to separate groups? Would the post-assessment's combined group score be the greatest and would it provide a broader data range? These questions highlight the need for further research to build upon the findings of this experiment. They also show that research is needed to explore the pros and cons of using digital tools to learn math or whether it should be used as a supplement to traditional teaching methods. Technology should be used to enhance and support traditional teaching methods instead of replacing them entirely. The idea is to integrate technology into the existing educational framework to enrich the learning experience (Newton et al., 2020). Also, it should be used effectively because the crucial factor is how teachers utilize this technology in their teaching (Zhang et al., 2020). For instance a tablet can be a powerful educational tool when used interactively by teachers or facilitators to engage students, explain concepts or provide personalized learning experiences. The study was mainly designed to find innovative and effective ways to use technology to enhance traditional teaching methods. This provides a richer and more diverse learning environment for students to help develop their STEM skills. The experiment at Clarksburg Elementary School has opened a door to a new perspective on how digital tools can be used to enhance students' math skills. However, walking through that door requires a deeper understanding and further exploration to ensure that the full potential of digital tools can be harnessed to benefit students in their mathematical journey.

## Conclusion

In conclusion, the data from this study strongly supports the integration of digital tools in mathematics education. The increased improvement percentages, higher engagement levels, immediate feedback and access to a broad array of resources indicate that digital tools can create a more conducive learning environment for mathematics. These findings offer compelling evidence for the potential benefits of incorporating digital tools into educational curriculum. This research contributes to a growing body of evidence that digital tools can enhance the educational process. As education professionals continue to explore the challenges and opportunities from the use of technology, studies like this one are imperative to guide the way forward.

# Appendix

**Pre-Test**

1. What is the probability of flipping a coin and it landing on heads?
a) 1/3
b) 1/2
c) 1/4
d) 1

2. If you roll a standard six-sided die, what is the probability of rolling a number greater than 4?
a) 1/6
b) 2/6
c) 1/3
d) 4/6

3. You have a spinner with 4 equal sections: Red, Blue, Green, and Yellow. What is the probability of landing on Green?
a) 1/4
b) 1/2
c) 1/3
d) 1/8

4. There are 3 red, 2 blue, and 1 green marble in a bag. If you pick one marble without looking, what is the probability of picking a blue marble?
a) 1/3
b) 1/2
c) 2/6
d) 1/6

5. If you roll a die and flip a coin at the same time, how many possible outcomes are there?
a) 12
b) 10
c) 8
d) 6

6. What is the probability of not landing on tails when flipping a coin?
a) 1
b) 1/2
c) 1/3
d) 1/4

7. Which event has a greater probability: Rolling a number less than 4 on a six-sided die or flipping a coin and getting heads?
a) Rolling a number less than 4
b) Flipping a coin and getting heads
c) Both are the same
d) Neither

8. If the odds of drawing a red marble from a bag are 3:2, how many red marbles are there compared to non-red marbles?
a) 1 red to 2 non-red
b) 3 red to 2 non-red
c) 2 red to 3 non-red
d) 3 red to 1 non-red

9. If you flip a coin 10 times and it lands on heads 7 times, what is the experimental probability of getting heads?
a) 1/2
b) 1/3
c) 7/10
d) 3/10

10. What is the probability of drawing a red card from a standard deck of cards?
a) 1/4
b) 1/2
c) 1/3
d) 1/6

# Post-Test

1. What is the chance of rolling a die and getting an even number?
a) 1/6
b) 2/6
c) 3/6
d) 4/6

2. If you pick a card from a deck of 52 cards, what is the probability of getting a heart?
a) 1/2
b) 1/3
c) 1/4
d) 1/13

3. You have a bag with 5 red, 3 blue, and 2 yellow marbles. What is the probability of drawing a yellow marble?
a) 1/5
b) 1/10
c) 1/2
d) 1/3

4. If you spin a spinner with 8 equal sections, what is the probability of landing on a number less than 5?
a) 3/8
b) 1/8
c) 4/8
d) 5/8

5. What is the chance of picking a red marble from a bag with 4 red, 3 blue, and 5 green marbles?
a) 1/12
b) 4/12
c) 7/12
d) 5/12

6. If you flip a coin twice, what is the probability of getting heads both times?
a) 1/2
b) 1/4
c) 3/4
d) 1

7. What is the likelihood of rolling a number greater than 2 on a six-sided die?
a) 1/2
b) 1/3
c) 4/6
d) 5/6

8. If you pick a marble from a bag with 2 red, 2 blue, and 6 green marbles, what is the probability of picking a red or blue marble?
a) 4/10
b) 6/10
c) 8/10
d) 10/10

9. If you roll two dice, what is the probability of the sum being 7?
a) 1/12
b) 1/6
c) 2/6
d) 5/6

10. What is the probability of not picking a face card from a standard deck of cards?
a) 4/52
b) 12/52
c) 40/52
d) 52/52